\documentclass[%
    aps,
    prd,
    10pt,
    nofootinbib,
    noeprint,
    amsmath,
    amssymb,
    eqsecnum
]{revtex4-2}

\usepackage[%
    colorlinks=true,
    allcolors=blue
]{hyperref}

\def\apss{Astrophys.\ Space\ Sci.}
\def\cqg{Classical\ Quant.\ Grav.}
\def\cmp{Commun.\ Math.\ Phys.}
\def\lrr{Living\ Rev.\ Relativity}
\def\mnras{Mon.\ Not.\ R.\ Astron.\ Soc.}
\def\prsa{Proc.\ R.\ Soc.\ A}

\begin{document}

\title{Perturbation theory for post-Newtonian neutron stars}

\author{Fabian Gittins}
\email{f.w.r.gittins@uu.nl}
\affiliation{Institute for Gravitational and Subatomic Physics (GRASP), Utrecht University, Princetonplein 1, 3584 CC Utrecht, Netherlands}
\affiliation{Nikhef, Science Park 105, 1098 XG Amsterdam, Netherlands}

\author{Nils Andersson}
\affiliation{Mathematical Sciences and STAG Research Centre, University of Southampton, Southampton SO17 1BJ, United Kingdom}

\author{Shanshan Yin}
\affiliation{Mathematical Sciences and STAG Research Centre, University of Southampton, Southampton SO17 1BJ, United Kingdom}

\date{\today}

\begin{abstract}
    Neutron stars are compact, relativistic bodies that host several extremes of modern physics. An exciting development in recent years has been the opportunity to probe this exotic physics by observing compact-binary coalescences using sensitive gravitational-wave and electromagnetic instruments. To maximise the science inferred from these measurements, we require models that accurately represent the physics. In this study, we consider the post-Newtonian approximation to general relativity for the modelling of neutron-star dynamics, with a particular view to model  dynamical tides at the late stages of binary inspiral. We develop the post-Newtonian perturbation equations for a non-rotating star and show that the perturbation problem is Hermitian and therefore derives from a fundamental Lagrangian. Establishing this Lagrangian system leads to a conserved symplectic product and canonical energy for the perturbations. We determine the orthogonality condition for the post-Newtonian oscillation modes, which in turn forms the foundation of a mode-sum representation often used for dynamical tides. Finally, we demonstrate that the perturbation formulation is unique.
\end{abstract}

\maketitle

\section{Introduction}

Neutron-star binaries produce cosmic fireworks in the Universe, as they gradually inspiral due to the emission of gravitational waves and eventually merge giving rise to powerful gamma-ray bursts and bright kilonovae. The breakthrough gravitational-wave observation GW170817 \cite{2017ApJ...848L..12A,2017ApJ...848L..13A,2017PhRvL.119p1101A}---along with the associated detections across the electromagnetic spectrum---began the enterprise of mining these remarkable events for information on high-density nuclear physics \cite{2018PhRvL.121p1101A}, astrophysics \cite{2017ApJ...850L..40A}, as well as cosmology \cite{2017Natur.551...85A}.

A key problem in the study of neutron-star binaries is how to describe their structure and gravitational-wave emission during the inspiral phase. For the early part of the inspiral, the orbit varies slowly and the tidal response of each star is well approximated as static. However, during the late inspiral, the orbital evolution sources a time-varying force that drives the oscillation modes of the neutron star, many of which may become resonantly excited. This is known as the dynamical tide.

In the absence of dissipation, it has been shown that the Newtonian mode equations are self-adjoint \cite{1964ApJ...139..664C}, implying that one can expand a generic perturbation of the star in terms of its natural vibrations \cite{1979RSPSA.368..389D,2001PhRvD..65b4001S}. The \textit{mode-sum} representation elegantly decouples the time-dependent sector of the problem into that of a forced harmonic oscillator and has been applied in the Newtonian tidal problem \cite{1994ApJ...426..688R,2020PhRvD.101h3001A}. However, such a decomposition is not formally possible in general relativity, since the theory is inherently dissipative due to the emission of gravitational radiation. This presents a problem as the neutron stars that we wish to model fall well outside the Newtonian regime. In the spirit of improving upon Newtonian calculations, we explore how post-Newtonian (pN) theory may be used to represent neutron-star dynamics. This is expected to lead to better models given that the internal energy contribution to the matter equation of state will be accounted for.

Formally, pN theory is a weak-field, slow-motion approximation to general relativity, which already serves as a vital tool in describing the orbital evolution of compact binaries \cite{2014grav.book.....P,2024LRR....27....4B}. In the approximation, one implicitly assumes the small parameter
\begin{equation}
    \lambda = \text{max} \left( \frac{U}{c^2}, \frac{v^2}{c^2}, \frac{p}{\rho c^2}, \frac{\Pi}{c^2} \right) \ll 1,
    \label{eq:Small}
\end{equation}
where $c$ is the speed of light, $U$ is the gravitational potential and $v^j$, $p$, $\rho$ and $\Pi$ are the three-velocity, pressure, mass density and specific internal energy, respectively, of the matter. In a slight abuse of notation, it is common to truncate the expressions to some $O(c^{-2n})$, which represents the $n$th pN order. In this paper, we will focus on the first pN order (1pN) and discard terms above $O(c^{-2})$.

It is worth noting that each of the quantities in Eq.~\eqref{eq:Small} is a separate measure. Indeed, one may be interested in a physical situation where the gravity is weak, but the velocities are relativistic. Or one may focus on the opposite regime. Neither of these situations would be consistent with the pN approximation, which assumes that \textit{all} of the parameters in Eq.~\eqref{eq:Small} are small.This approximation is perfectly appropriate in the far-field, vacuum exterior of a neutron star, where the expansion is well-controlled and widely used to describe orbital dynamics (provided the motion is suitably slow). However, it inevitably breaks down closer to the star and inside its interior. Due to their compactness, neutron stars host strong gravitational fields, such that there is a point in the approach from the far field when it is no longer valid to assume $U / c^2 \ll 1$. Inside the star, the fluid exhibits large $p$, $\rho$ and $\Pi$, making the formal pN expansion problematic for neutron-star hydrodynamics. However, our goal is modest: We aim to use pN theory to provide a more accurate description of neutron-star dynamics than Newtonian gravity. In particular, we consider how oscillations computed in the pN approximation form a complete basis in terms of which the tidal response of a neutron star can be decomposed. The presence of a companion star will raise a tide on the neutron star, which at leading Newtonian order manifests as an additional gravitational potential. While pN corrections to this tidal field are expected, determining their precise form requires a more detailed analysis, which we relegate to future work. The ultimate description should obviously involve full general relativity, but as there are challenges that prevent progress in this direction \cite{2024PhRvD.109f4004P,2024PhRvD.109j4064H}, the pN models we outline here should prove useful.

In a formal pN expansion, it is well known that dissipation due to gravitational waves manifests at 2.5pN through the Burke-Thorne radiation-reaction force \cite{1969PhDT.......152B,1969ApJ...158..997T,1970ApJ...160..153C}. Therefore, in principle, one should be able to work in pN theory up to 2pN and the equations of motion should remain self-adjoint, facilitating a mode-sum expansion. Indeed, the self-adjoint nature of the pN equations was explored by \citet{1965ApJ...142.1519C} and we develop upon this work in this paper.

\section{Post-Newtonian hydrostatic equilibrium}

\subsection{Standard formulation}

We start by considering the pN equations that describe the structure of a non-rotating, self-gravitating fluid (see, \textit{e.g.}, Sec.~8.4 of Ref.~\cite{2014grav.book.....P}  and the discussion in Ref.~\cite{2023CQGra..40b5016A}). This will form the background on which we construct the perturbations later. For a static fluid, the pN Euler equation can be written as
\begin{equation}
    \left[ 1 - \frac{1}{c^2} \left( U + \Pi + \frac{p}{\rho^*} \right) \right] \frac{dp}{dr} = \rho^* \left( 1 - \frac{1}{c^2} 4 U \right) \frac{dU}{dr} + \frac{1}{c^2} \rho^* \frac{d\psi}{dr},
    \label{eq:Euler}
\end{equation}
where $r$ is the isotropic radial coordinate, the potentials $U$ and $\psi$ are given by the Poisson equations
\begin{subequations}
\begin{gather}
    \nabla^2 U = - 4 \pi G \rho^*, \label{eq:U} \\
    \nabla^2 \psi = - 4 \pi G \rho^* \left( - U + \Pi + 3 \frac{p}{\rho^*} \right), \label{eq:psi}
\end{gather}
\end{subequations}
$G$ is Newton's gravitational constant, $\nabla^2$ represents the flat-space Laplacian and
\begin{equation}
    \rho^* = \rho \left( 1 + \frac{1}{c^2} 3 U \right)
    \label{eq:RescaledMassDensity}
\end{equation}
is a re-scaled mass density. For neutron stars, it is useful to decompose the fluid mass density as $\rho = m_\text{b} n_\text{b}$, where $n_\text{b}$ is the baryon-number density and $m_\text{b}$ denotes the mass per baryon.

Reflecting on the form of these equations, we make the following observations. First, we are evidently not---noting the explicit use of $\rho^*$---dealing with a strict order-by-order expansion in terms of $c^{-2}$. Rather, the form of the equations is informed by the physics intuition. In particular, conservation of baryons (which we discuss later) requires that
\begin{equation}
    \partial_t \rho^* + \partial_j (\rho^* v^j) = 0,
    \label{eq:BaryonMassConservation}
\end{equation}
where $\partial_t$ and $\partial_j$ are the partial time and spatial derivatives, respectively. It is easy to see that the use of $\rho^*$ ensures that the baryon conservation law is satisfied ``at all orders'' \cite{2014grav.book.....P}. This is obviously convenient. Second, it is evident from the manifestation of the scalar potential $\psi$ at 1pN in the Euler equation~\eqref{eq:Euler} that the specific internal energy $\Pi$ has now entered the hydrodynamical description. This is notably not the case in Newtonian gravity, and is a relativistic contribution. However, this means that the assumption $\Pi / c^2 \ll 1$ has been made. This makes sense for non-relativistic stars, but it may not be a valid assumption for neutron stars. Of course, one may reasonably argue that neutron stars are not in the weak-field regime anyway so the entire discussion is somewhat moot. Be that as it may, we want to explore to what extent we can develop viable models for neutron-star dynamics and compare their performance---\textit{e.g.}, in terms of the relevant mode-oscillation frequencies and the associated tidal response---to alternative approaches \cite{pn_modes}.

Moving on, the variable $\rho^*$ determines the star's baryon mass $M_\text{b}$ through
\begin{subequations}\label{eqs:Structure}
\begin{equation}
    \frac{dM_\text{b}}{dr} = 4 \pi r^2 \rho^*.
    \label{eq:dMb_dr}
\end{equation}
Combining the Poisson equation~\eqref{eq:U} with Eq.~\eqref{eq:dMb_dr}, a trivial integration from the stellar centre, where $dU / dr$ must be regular and $M_\text{b}$ vanishes, up to some radius $r$ provides
\begin{equation}
    \frac{dU}{dr} = - \frac{G M_\text{b}}{r^2}.
\end{equation}
Truncating the equation of hydrostatic equilibrium~\eqref{eq:Euler} to leading order, we arrive at the usual Newtonian equations of stellar structure. It is interesting to note that Eq.~\eqref{eq:dMb_dr} involves no approximation. This is because the re-scaled density $\rho^*$ involves the metric determinant [see Eq.~\eqref{eq:RescaledMassDensity2} below] and implicitly integrates with respect to the proper volume of the spacetime. The baryon mass $M_\text{b}$ is the conserved quantity implied by the conservation law~\eqref{eq:BaryonMassConservation}, and this remains the case in full general relativity.

Motivated by the Newtonian expressions, we define another mass function $\mathcal{N}$ by
\begin{equation}
    \frac{d\mathcal{N}}{dr} = 4 \pi r^2 \rho^* \left( - U + \Pi + 3 \frac{p}{\rho^*} \right),
\end{equation}
so---invoking similar regularity arguments as above---the potential $\psi$ satisfies
\begin{equation}
    \frac{d\psi}{dr} = - \frac{G \mathcal{N}}{r^2}.
\end{equation}
With these preliminaries, we can express hydrostatic equilibrium~\eqref{eq:Euler} as
\begin{equation}
    \frac{dp}{dr} = - \frac{G \rho^*}{r^2} \left\{ M_\text{b} + \frac{1}{c^2} \left[ \left( - 3 U + \Pi + \frac{p}{\rho^*} \right) M_\text{b} + \mathcal{N} \right] \right\}
\end{equation}
\end{subequations}
and Eqs.~\eqref{eqs:Structure} characterise the structure of the star up to 1pN. In a similar fashion to how $U$ decouples from the Newtonian system, the value of $\psi$ does not manifest itself at 1pN; although its derivative does. It is also useful to note that the star possesses gravitational mass $M = M_\text{b} + \mathcal{N} / c^2$. As usual, we require  additional information that describes the microscopic interactions in the stellar fluid---an \textit{equation of state}---to close the system. In particular, we need relationships between the variables $p$, $\Pi$ and $\rho^*$. We will resist the temptation to assume the functions $p = p(\rho^*)$ and $\Pi = \Pi(\rho^*)$ for reasons that will be explained.

For illustration purposes, let us assume that the neutron star is cold and composed of neutrons, protons and electrons. Under these assumptions, the first law of thermodynamics can be expressed as
\begin{equation}
    d\varepsilon = \frac{\varepsilon + p}{n_\text{b}} \, dn_\text{b} + n_\text{b} \mu_\Delta \, dY_\text{e},
    \label{eq:FirstLaw}
\end{equation}
where $\varepsilon$ is the total energy density of the fluid, $Y_\text{e}$ is the electron fraction and $\mu_\Delta$ is the chemical potential that determines the balance in nuclear reactions. In general, this implies a fundamental potential of the form $\varepsilon = \varepsilon(n_\text{b}, Y_\text{e})$, or equivalently $\varepsilon = \varepsilon(\rho, Y_\text{e})$. When the star is in equilibrium $\mu_\Delta = 0$, the material is barotropic and the equation of state is $\varepsilon = \varepsilon(\rho)$. From the equation of state, one can derive the pressure via
\begin{equation}
    p(\rho, Y_\text{e}) = \rho^2 \left[ \frac{\partial (\varepsilon / \rho)}{\partial \rho} \right]_{Y_\text{e}},
\end{equation}
which reduces to the function $p = p(\rho)$ when the matter is in equilibrium. For neutron-star matter, this information is usually provided in tabulated form.

Next, we can use the exact relation
\begin{equation}
    \varepsilon = \rho c^2 \left( 1 + \frac{1}{c^2} \Pi \right)
    \label{eq:varepsilon}
\end{equation}
to calculate $\Pi$ from the equation-of-state model. This is sufficient to determine $\Pi = \Pi(\rho)$. We can then use Eq.~\eqref{eq:RescaledMassDensity} to derive relationships in terms of $\rho^*$ and calculate the pN stellar structure. Along with the standard boundary conditions---zero mass at the centre, potentials matching smoothly to their exterior (near-zone) values and vanishing pressure at the surface---we have all the information we need to progress.

In Ref.~\cite{2023CQGra..40b5016A}, we discussed different models accurate to first pN order, involving alternative choices of which  higher order terms to include or omit. This demonstrated that---perhaps not surprisingly for neutron-star densities and pressures---the inclusion of different sets of higher order pN terms can make a notable difference in the ``accuracy'' of the static neutron-star models obtained for the same realistic equation of state. The freedom of choice associated with this is, at least to some extent, disturbing because it means that any calculation we carry out will inevitably involve some level of ``black magic''. This is not an uncommon feature of approximation theory, but it is still nevertheless somewhat uncomfortable. Evidently, pN modelling involves making choices. The only alternative would be to carry out the calculation at a strict order-by-order basis. While this is not the usual strategy, it is worth asking to what extent this could be a viable approach.

\subsection{Order-by-order expansion}

The kind of problem we are dealing with typically lends itself to an order-by-order approach, where one assumes a power-series expansion in the small parameter for all variables of the problem and then iteratively calculates the variables at successive powers in the small parameter. This approach is commonly used in slow-rotation approximations \cite{1978trs..book.....T,1981Ap&SS..78..483S}. However, the order-by-order approach is generally not used in modern pN theory. (The notable exceptions to this are studies by Cutler and Lindblom~\cite{1991ApJ...374..248C,1992ApJ...385..630C} and Futamase and Schutz~\cite{1983PhRvD..28.2363F,1983PhRvD..28.2373F}.) Here, we will examine why this is the case for the hydrodynamics.

The equilibrium equations were written down with the attitude that we can freely keep some higher order pN contributions in the calculation. This should (obviously) be fine as long as the pN corrections are small, but we know the higher order terms make a significant difference for neutron-star models. This was clearly demonstrated in Ref.~\cite{2023CQGra..40b5016A}. As a step towards understanding the issues we face when we turn to the perturbation problem, let us briefly explore what happens if we try to strictly solve for the background star order by order in the pN expansion. The appeal of such an expansion would be that it deals with each order systematically and removes the arbitrariness in pN hydrodynamics explored in Ref.~\cite{2023CQGra..40b5016A}.

For this discussion, it will be sufficient to focus on the thermodynamical aspects of the stellar fluid. We begin by expanding our variables as
\begin{equation}
    p = p_0 + \frac{1}{c^2} p_1, \qquad \rho = \rho_0 + \frac{1}{c^2} \rho_1, \qquad \varepsilon = c^2 \varepsilon_0 + \varepsilon_1, \qquad \Pi = \Pi_0 + O(c^{-2}).
\end{equation}
Here, we identify terms with subscript $0$ as the leading-order, Newtonian variables, while the subscript $1$ denotes the pN corrections. By Eq.~\eqref{eq:varepsilon}, we can deduce that
\begin{equation}
    \varepsilon_0 = \rho_0, \qquad \varepsilon_1 = \rho_1 + \rho_0 \Pi_0.
\end{equation}
With these identifications, we are in a position to consider the equation of state.

From the perspective of Newtonian fluid mechanics, it seems natural to work with the functions $p = p(\rho)$ and $\Pi = \Pi(\rho)$. Indeed, expanding these functions leads to the relations
\begin{equation}
    p_0 = p(\rho = \rho_0), \qquad p_1 = \left. \frac{dp}{d\rho} \right|_{\rho = \rho_0} \rho_1, \qquad \Pi_0 = \Pi(\rho = \rho_0).
\end{equation}
The leading term that determines $p_0$ from $\rho_0$ is precisely the same relation that closes the Newtonian structure equations. The additional relations for $p_1$ and $\Pi_0$ manifest at 1pN.

Alternatively, for fully relativistic neutron stars, it is more natural to consider the function $p = p(\varepsilon)$, which may also be deduced from the equation of state. (This is the approach taken in Ref.~\cite{1991ApJ...374..248C}.) If we instead expand this function, along with $\Pi = \Pi(\varepsilon)$, we find
\begin{equation}
    p_0 = p(\varepsilon = c^2 \rho_0), \qquad p_1 = c^2 \left. \frac{dp}{d\varepsilon} \right|_{\varepsilon = c^2 \rho_0} (\rho_1 + \rho_0 \Pi_0), \qquad \Pi_0 = \Pi(\varepsilon = c^2 \rho_0).
\end{equation}
Evidently, the values obtained for (say) $p_0$ are different in these two approaches; a realistic equation of state will not, in general, yield the same pressures at $\rho = \rho_0$ and $\varepsilon = c^2 \rho_0$. (Although, at densities and pressures where the pN approximation is valid, these expressions will be the same up to pN corrections.) Therefore, a Newtonian neutron star that is generated at leading order in the expansion will possess different properties depending on the adopted approach. It is not clear which strategy one should favour and this ambiguity makes an order-by-order expansion for the hydrodynamics unattractive.

Furthermore, when it comes to the matter description, there is a fundamental axiom that calls into question any expansion that involves, or is related to, the gravitational field. Due to the equivalence principle, the laws of thermodynamics should have the same form in all frames of reference. Therefore, we are always free to examine a locally flat region of the spacetime and construct the first law in the form of \eqref{eq:FirstLaw}, where the quantities are as measured by a local, inertial observer. This means that the equation of state $\varepsilon = \varepsilon(\rho)$ cannot be cognizant of the spacetime geometry and thus any expansion that is related to the varying curvature is dubious from this perspective. Indeed, it is for this reason that we withheld from introducing the functions $p = p(\rho^*)$ and $\Pi = \Pi(\rho^*)$ to close the pN equations of stellar structure~\eqref{eqs:Structure}; $\rho^*$ is not a thermodynamical variable since it explicitly depends on the gravitational field [\textit{cf.}, Eq.~\eqref{eq:RescaledMassDensity}].

\section{Towards post-Newtonian hydrodynamics}

Having argued against an order-by-order expansion for the pN hydrodynamics, we will now lay out the steps that lead to the usual formulation, where the equations are truncated at $O(c^{-2})$ (implicitly retaining some higher order terms \cite{2023CQGra..40b5016A}). As we build towards describing neutron-star dynamics using the pN approximation, we will relax the constraint that the fluid  is static.

\subsection{General-relativistic hydrodynamics}

We take the fully relativistic equations as the natural starting point. Treating the stellar material as a perfect fluid, we then have the stress-energy tensor
\begin{equation}
    T^{a b} = \frac{1}{c^2} \varepsilon \, u^a u^b + p \perp^{a b},
\end{equation}
where the fluid four-velocity $u^a$ is normalised in such a way that $u_a u^a = - c^2$ and the orthogonal projection is defined by 
\begin{equation}
    \perp^{a b} = g^{a b} + \frac{1}{c^2} u^a u^b,
\end{equation}
where $g^{a b}$ is the (inverse) metric. We obtain the equations of fluid dynamics by evaluating
\begin{equation}
    \nabla_b T^{a b} = 0,
    \label{eq:Conservation}
\end{equation}
where $\nabla_a$ is the covariant derivative.

Equation~\eqref{eq:Conservation} leads to two conservation laws. The first is obtained by projecting Eq.~\eqref{eq:Conservation} along $u_a$, where we obtain the law of energy conservation
\begin{equation}
    \partial_b (\sqrt{-g} \varepsilon u^b) + p \partial_b (\sqrt{- g} u^b) = 0,
    \label{eq:EnergyConservation}
\end{equation}
where $g$ is the metric determinant. The second relation comes from projecting Eq.~\eqref{eq:Conservation} with $\perp_a^c$. Thus, we arrive at momentum conservation (the relativistic Euler equation)
\begin{equation}
    \perp^{c b} \partial_b p + \frac{1}{c^2} (\varepsilon + p) u^b \left( \partial_b u^c + \Gamma^c_{a b} u^a \right) = 0,
    \label{eq:MomentumConservation}
\end{equation}
where $\Gamma^a_{b c}$ is a Christoffel symbol.

At this point, it is instructive to demonstrate the consistency of the thermodynamics with the conservation of baryon number. The relevant conservation law takes the form
\begin{equation}
    \partial_a \left( \sqrt{-g} n_\text{b} u^a \right) = 0.
    \label{eq:BaryonConservation}
\end{equation}
The same relation follows from combining energy conservation~\eqref{eq:EnergyConservation} with the first law of thermodynamics~\eqref{eq:FirstLaw} and assuming that%
\footnote{The assumption of Eq.~\eqref{eq:Assumption} follows from either chemical equilibrium $\mu_\Delta = 0$ or the composition being advected with the fluid flow $u^a \partial_a Y_\text{e} = 0$. For an equilibrium neutron star, the former is perfectly appropriate. However, in dynamical contexts, the nuclear reactions are typically too slow to maintain chemical equilibrium and it may instead be assumed that $u^a \partial_a Y_\text{e} = 0$. That is, the matter composition is frozen and advected along with the fluid.} 
\begin{equation}
  \mu_\Delta u^a \partial_a Y_\text{e} = 0.
  \label{eq:Assumption}
\end{equation}
This is important if we want to build models based on realistic microphysics.

Before we move on, we will show how Eq.~\eqref{eq:BaryonMassConservation} is the statement of baryon-number conservation. First, we decompose the fluid four-velocity as
\begin{equation}
    u^t = \gamma c, \qquad u^j = \gamma v^j,
\end{equation}
with factor $\gamma = u^t / c$ and three-velocity $v^j$. Then multiplying Eq.~\eqref{eq:BaryonConservation} by the baryon mass $m_\text{b}$ leads to Eq.~\eqref{eq:BaryonMassConservation}, where we identify the re-scaled mass density
\begin{equation}
    \rho^* = \sqrt{-g} \gamma \rho.
    \label{eq:RescaledMassDensity2}
\end{equation}
Hence, Eq.~\eqref{eq:BaryonMassConservation} is the baryon-number conservation law. The argument also explains why it is natural to work with $\rho^*$ in the pN problem.

\subsection{Introducing the weak-field, slow-motion assumptions}

As a first step towards what may be meaningfully referred to as a pN model, let us assume that we are dealing with a weak gravitational field and slow fluid motion. We can make this notion precise by assuming the standard (near-zone) pN metric (in Cartesian coordinates) \cite{2014grav.book.....P}
\begin{subequations}\label{eqs:Metric}
\begin{align}
    g_{t t} &= - 1 + \frac{1}{c^2} 2 U + \frac{1}{c^4} 2 (\Psi - U^2) + O(c^{-6}), \\
    g_{t j} &= - \frac{1}{c^3} 4 U_j + O(c^{-5}), \\
    g_{j k} &= \delta_{j k} \left( 1 + \frac{1}{c^2} 2 U \right) + O(c^{-4}),
\end{align}
\end{subequations}
where $\delta_{j k}$ is the Kronecker delta and
\begin{equation}
    \Psi = \psi + \frac{1}{2} \partial_t^2 X.
\end{equation}
The potential $U$ satisfies Eq.~\eqref{eq:U}, which we reproduce here:
\begin{subequations}\label{eqs:Potentials}
\begin{equation}
    \nabla^2 U = - 4 \pi G \rho^*.
\end{equation}
The motion of the fluid sources the vector potential $U_j$ through
\begin{equation}
    \nabla^2 U_j = - 4 \pi G \rho^* v_j
\end{equation}
and also affects $\psi$ by modifying Eq.~\eqref{eq:psi} to 
\begin{equation}
   \nabla^2 \psi = - 4 \pi G \rho^* \left( \frac{3}{2} v^2 - U + \Pi + 3 \frac{p}{\rho^*} \right).
\end{equation}
The superpotential $X$ is a solution of
\begin{equation}
    \nabla^2 X = 2 U.
    \label{eq:X}
\end{equation}
\end{subequations}
Finally, it is relevant to record that the harmonic gauge condition has been assumed, leading to
\begin{equation}
    \partial_t U + \partial_j U^j = 0.
\end{equation}
It is easy to show that this relation is consistent with baryon conservation~\eqref{eq:BaryonMassConservation}.

With this set-up, we consider the momentum equation~\eqref{eq:MomentumConservation}. After some manipulation, we arrive at the pN Euler equation
\begin{equation}
\begin{split}
    \rho^* \frac{dv_j}{dt} = - \partial_j p + \rho^* \partial_j U &+ \frac{1}{c^2} \bigg\{ \bigg( \frac{1}{2} v^2 + \Pi + U + \frac{p}{\rho^*} \bigg) \partial_j p - v_j \partial_t p \\
    &+ \rho^* \big[ (v^2 - 4 U) \partial_j U - v_j (3 \partial_t U + 4 v^k \partial_k U) 
    + 4 \partial_t U_j + 4 v^k (\partial_k U_j - \partial_j U_k) + \partial_j \Psi \big] \bigg\}.
\end{split}
    \label{eq:EulerFull}
\end{equation}
In the static limit, this reduces to the equation of hydrostatic balance~\eqref{eq:Euler}.

Before we move on to consider pN perturbations, let us briefly discuss the solutions to Eqs.~\eqref{eqs:Potentials}. This gives us an opportunity to lay out the domain of integration, which will be important later on. The Poisson equations~\eqref{eqs:Potentials} admit the following Green's function solutions:
\begin{subequations}
\begin{gather}
    U = G \int \frac{\rho^{* \prime}}{|x^j - x^{\prime j}|} \, d^3 x', \\
    U_j = G \int \frac{\rho^{* \prime} v'_j}{|x^k - x^{\prime k}|} \, d^3 x', \\
    \psi = G \int \frac{\rho^{* \prime}}{|x^j - x^{\prime j}|} \left( \frac{3}{2} v^{\prime 2} - U' + \Pi' + 3 \frac{p'}{\rho^{* \prime}} \right) \, d^3 x',
\end{gather}
where $d^3 x$ is the Cartesian three-volume element and we use primes to indicate that a variable depends on $x^{\prime j}$, such as $\rho^{* \prime} = \rho^*(t, x^{\prime j})$, whereas unprimed variables depend on $x^j$. It is important to observe that the integrals are naturally truncated to the volume occupied by the matter distribution since the integrands vanish in the outside vacuum. However, although the fluid variables only have compact support on the star, the potentials still extend into the star's vacuum exterior.%
\footnote{We need to be mindful of the constraint on how far away from the matter distribution the pN metric~\eqref{eqs:Metric} (and therefore the potentials $U$, $U_j$, $\psi$ and $X$) applies. Strictly, it is only valid within the \textit{near zone}. This is the region of space enveloped by the characteristic wavelength of radiation, whereas the region beyond this is known as the \textit{wave zone}, where the pN approximation breaks down and retardation effects become relevant.}
For this reason, the solution to Eq.~\eqref{eq:X} is somewhat more subtle. The solution is given by
\begin{equation}
    X = G \int \rho^{* \prime} |x^j - x^{\prime j}| \, d^3 x',
\end{equation}
\end{subequations}
which is also automatically tied to the stellar volume. We note that this is, in fact, a definition as we ignore the homogeneous solution that also satisfies Eq.~\eqref{eq:X}. (For more detail on this issue, see Ref.~\cite{2014grav.book.....P}.)

\section{Post-Newtonian perturbations}

Now, we consider linear perturbations of a spherically symmetric pN star described by Eqs.~\eqref{eqs:Structure}. We follow the Lagrangian perturbation framework described in Ref.~\cite{1978ApJ...221..937F} and adapt it for pN hydrodynamics.

In the Newtonian theory, stellar perturbations can be fully described by the Lagrangian displacement vector $\xi^j$, which connects
the fluid elements of the star in the perturbed configuration with their original, equilibrium position. We will see shortly that, at first pN order, this remains the fundamental variable governing the perturbation dynamics. However, we note that in full general relativity, the situation is more intricate: in addition to the displacement vector, one must also account for perturbations of the spacetime metric itself \cite{1978CMaPh..62..247F}.

It is useful to introduce the following perturbation quantities. We denote the Lagrangian change of a quantity $Q$ (which can be scalar, vectorial or tensorial) by $\Delta Q$, whereas the Eulerian variation is $\delta Q$. The two are related by
\begin{equation}
    \Delta Q = \delta Q + \mathcal{L}_\xi Q,
\end{equation}
where $\mathcal{L}_\xi$ is the Lie derivative along
$\xi^j$.

The Lagrangian perturbation of the fluid velocity is
\begin{equation}
    \Delta v^j = \partial_t \xi^j.
\end{equation}
Since the background is static, the Eulerian change is identical, $\delta v^j = \partial_t \xi^j$. Conservation of baryon mass during the perturbation leads to
\begin{equation}
    \Delta \rho^* = - \rho^* \partial_j \xi^j.
\end{equation}
We treat the composition of the star as frozen during a perturbation, such that $\Delta Y_\text{e} = 0$. This amounts to assuming that all nuclear reactions are slow enough (compared to the dynamics) that they can be ignored. This assumption is expected to be relevant for the late-inspiral phase of compact-binary systems involving neutron stars. In effect, the thermodynamical function $p = p(\rho, Y_\text{e})$ implies that
\begin{equation}
    \frac{\Delta p}{p} = \Gamma_1 \frac{\Delta \rho}{\rho} = \Gamma_1 \left( \frac{\Delta \rho^*}{\rho^*} - \frac{1}{c^2} 3 \Delta U \right),
    \label{eq:Deltap}
\end{equation}
where $\Gamma_1 = (\partial p / \partial \rho)_{Y_\text{e}}$ is the adiabatic index.%
\footnote{The perturbation equation~\eqref{eq:Deltap} may also be applied to other stellar models, not just neutron stars (see, \textit{e.g.}, Ref.~\cite{1978ApJ...221..937F}). However, the assumed, underlying physics is different. Indeed, for some stars, the nuclear reactions are slow, but the matter has finite temperature. Under these conditions, we instead have $p = p(\rho, s)$ and $\Gamma_1 = (\partial p / \partial \rho)_s$, where $s$ is the entropy per baryon. This amounts to assuming adiabatic perturbations $\Delta s = 0$. Here, we assumed zero temperature and therefore excluded the temperature term in the first law~\eqref{eq:FirstLaw}.}
For the potentials, it is convenient to use Eulerian perturbations, where we have
\begin{subequations}\label{eqs:PerturbedPotentials}
\begin{gather}
    \delta U = G \int \frac{\delta \rho^{* \prime}}{|x^j - x^{\prime j}|} \, d^3 x', \label{eq:deltaU} \\
    \delta U_j = G \int \frac{\rho^{* \prime} \delta v'_j}{|x^k - x^{\prime k}|} \, d^3 x', \\
    \delta \psi = G \int \frac{1}{|x^j - x^{\prime j}|} \left[ \delta \rho^{* \prime} (- U' + \Pi') + \rho^{* \prime} \left( - \delta U' + \delta \Pi' + 3 \frac{\delta p'}{\rho^{* \prime}} \right) \right] \, d^3 x', \label{eq:deltapsi} \\
    \delta X = G \int \delta \rho^{* \prime} |x^j - x^{\prime j}| \, d^3 x'.
\end{gather}
For the perturbation problem, it is useful to introduce the new vector potential
\begin{equation}
    \delta V_j = G \int \frac{\rho^{* \prime} \xi'_j}{|x^k - x^{\prime k}|} \, d^3 x',
\end{equation}
\end{subequations}
which is related to the previous vector potential by $\delta U_j = \partial_t \delta V_j$. Additionally, it will be helpful to note the identity
\begin{equation}
    \partial_j \delta X + \delta V_j = G \int \rho^{* \prime} \xi^{\prime k} \frac{(x_j - x^\prime_j) (x_k - x^\prime_k)}{|x^l - x^{\prime l}|^3} \, d^3 x'.
\end{equation}

Finally, we linearise the Euler equation~\eqref{eq:EulerFull} to arrive at the equation of motion
\begin{subequations}\label{eqs:PerturbedEuler}
\begin{equation}
    0 = A_{j k} \partial_t^2 \xi^k + C_{j k} \xi^k,
\end{equation}
where the operators $A_{j k}$ and $C_{j k}$ are given by
\begin{align}
    A_{j k} \partial_t^2 \xi^k &= \rho^* \left\{ \left[ 1 + \frac{1}{c^2} \left( 3 U + \Pi + \frac{p}{\rho^*} \right) \right] \partial_t^2 \xi_j - \frac{1}{c^2} \left( 4 \partial_t^2 \delta V_j + \frac{1}{2} \partial_t^2 \partial_j \delta X \right) \right\}, \\
\begin{split}
    C_{j k} \xi^k &= \left( 1 + \frac{1}{c^2} 2 U \right) \partial_j \delta p - \rho^* \left[ 1 + \frac{1}{c^2} \left( - U + \Pi + \frac{p}{\rho^*} \right) \right] \partial_j \delta U - \frac{1}{c^2} \left( - 3 \delta U + \delta \Pi + \frac{\delta p}{\rho^*} \right) \partial_j p \\
    &\qquad- \frac{\delta \rho^*}{\rho^*} \left[ 1 + \frac{1}{c^2} \left( 2 U - \frac{p}{\rho^*} \right) \right] \partial_j p - \frac{1}{c^2} \rho^* \partial_j \delta \psi.
\end{split}
\end{align}
\end{subequations}
The perturbations are all fundamentally related to the fluid displacement $\xi^j$ through the expressions we have provided. We note that the equations reduce to the standard Newtonian perturbation equations when terms of $O(c^{-2})$ are discarded (\textit{cf.}, Ref.~\cite{1978ApJ...221..937F}). For reasons that will become clear in the following, we have multiplied the linearised Euler equation by $[1 + (3 U + \Pi + p / \rho^*) / c^2]$ to reproduce Eq.~(5) in Ref.~\cite{1965ApJ...142.1519C}.%
\footnote{A notable difference with our formulation and that of Ref.~\cite{1965ApJ...142.1519C} is that we choose to define the potentials in terms of $\rho^*$ instead of $\rho$, which means that they formally differ at first pN order.}
Indeed, there are many possible versions of $A_{j k}$ and $C_{j k}$, since we are always free to multiply the equation of motion by a pN term, while factoring out the higher order terms, and the result should capture the same physics (up to 2pN corrections). However, we will soon see how the precise form of the equation of motion plays a pivotal role in its symmetry properties. This is relevant as it will allow us to identify a unique definition of the Lagrangian for the perturbed system, along with the canonical energy and an orthogonality condition for oscillation modes.

\subsection{Symmetries}

To study the symmetry of the operators, we introduce the inner product
\begin{equation}
    \langle \eta, \xi \rangle = \int \bar{\eta}^j \xi_j \, d^3 x
    \label{eq:InnerProduct}
\end{equation}
between two vector solutions $\xi^j$ and $\eta^j$ to the equation of motion~\eqref{eqs:PerturbedEuler}, where the bar denotes complex conjugation \cite{1978ApJ...221..937F}. Since the displacements only have compact support on the star, the integration domain is evidently tethered to the matter distribution. This is identical to the linearised potentials~\eqref{eqs:PerturbedPotentials} and will be important when we consult the divergence theorem later.

In defining the inner product~\eqref{eq:InnerProduct}, we have elected to use the Cartesian three-volume element $d^3 x$. It is common to instead use the proper three-volume element $dV = \sqrt{- g} \, d^3 x$ (see, \textit{e.g.}, Ref.~\cite{2021MNRAS.506.2985K}). Another convention that exists in the literature is to define the inner product with the mass density $\rho$ \cite{1977ApJ...213..183P}. Favouring one convention over another does not affect the results, however, the operators that are Hermitian with respect to the chosen inner product will simply differ by factors of $\sqrt{- g}$ and $\rho$. Under the inner product, the quantities, such as energy and angular momentum, will be identical.

It was shown by \citet{1965ApJ...142.1519C} that the pN perturbation problem in the form of \eqref{eqs:PerturbedEuler} is self-adjoint. This is an important result since it implies that mode solutions to the linearised equation of motion~\eqref{eqs:PerturbedEuler} form a complete basis with which a generic perturbation of the star can be decomposed. To better understand this (and later expand on this result to explore the fundamental Lagrangian system), we will go through this exercise now using the modern formulation of the pN equations involving $\rho^*$. It is reassuring to see that the two formulations are equivalent in this sense (up to higher order corrections).

It is trivial to verify that the operator $A_{j k}$ is Hermitian with respect to the inner product~\eqref{eq:InnerProduct}:
\begin{equation}
\begin{split}
    \langle \eta, A \xi \rangle &= \int \bar{\eta}^j A_{j k} \xi^k \, d^3 x \\
    &= \int \bar{\eta}^j \rho^* \left\{ \left[ 1 + \frac{1}{c^2} \left( 3 U + \Pi + \frac{p}{\rho^*} \right) \right] \xi_j - \frac{1}{c^2} \left( 4 \delta_\xi V_j + \frac{1}{2} \partial_j \delta_\xi X \right) \right\} \, d^3 x \\
    &= \int \rho^* \left[ 1 + \frac{1}{c^2} \left( 3 U + \Pi + \frac{p}{\rho^*} \right) \right] \bar{\eta}^j \xi_j \, d^3 x - \frac{1}{c^2} \frac{7 G}{2} \int \int \frac{\rho^* \bar{\eta}^j \rho^{* \prime} \xi_j'}{|x^k - x^{\prime k}|} \, d^3 x' \, d^3 x \\
    &\quad- \frac{1}{c^2} \frac{G}{2} \int \int \rho^* \bar{\eta}^j \rho^{* \prime} \xi^{\prime k} \frac{(x_j - x^\prime_j) (x_k - x^\prime_k)}{|x^l - x^{\prime l}|^3} \, d^3 x' \, d^3 x \\
    &= \overline{\langle \xi, A \eta \rangle},
\end{split}
\end{equation}
where the notation $\delta_\xi$ denotes that the perturbation is sourced by the displacement $\xi^j$. We use $\delta_\eta$ to symbolise solutions corresponding to $\eta^j$. To arrive at the result, we have used the fact that the background quantities are all real-valued.

The analysis of $C_{j k}$ is more involved. In order to demonstrate that it is also Hermitian, we must evaluate
\begin{equation}
\begin{split}
    \langle \eta, C \xi \rangle &= \int \bar{\eta}^j C_{j k} \xi^k \, d^3 x \\
    &= \int \bar{\eta}^j \bigg\{ \left( 1 + \frac{1}{c^2} 2 U \right) \partial_j \delta_\xi p - \rho^* \left[ 1 + \frac{1}{c^2} \left( - U + \Pi + \frac{p}{\rho^*} \right) \right] \partial_j \delta_\xi U - \frac{1}{c^2} \left( - 3 \delta_\xi U + \delta_\xi \Pi + \frac{\delta_\xi p}{\rho^*} \right) \partial_j p \\
    &\qquad\qquad- \frac{\delta_\xi \rho^*}{\rho^*} \left[ 1 + \frac{1}{c^2} \left( 2 U - \frac{p}{\rho^*} \right) \right] \partial_j p - \frac{1}{c^2} \rho^* \partial_j \delta_\xi \psi \bigg\} \, d^3 x.
\end{split}
\end{equation}
We will go through this term by term.

First, we consider
\begin{equation}
\begin{split}
    I_1 &= \int \bar{\eta}^j \left[ \left( 1 + \frac{1}{c^2} 2 U \right) \partial_j \delta_\xi p - \frac{1}{c^2} \frac{\delta_\xi p}{\rho^*} \partial_j p \right] \, d^3 x \\
    &= \int \partial_j \left[ \bar{\eta}^j \left( 1 + \frac{1}{c^2} 2 U \right) \delta_\xi p \right] \, d^3 x - \int \left[ \partial_j \bar{\eta}^j \left( 1 + \frac{1}{c^2} 2 U \right) + \frac{1}{c^2} 3 \frac{1}{\rho^*} \bar{\eta}^j \partial_j p \right] \delta_\xi p \, d^3 x.
\end{split}
\end{equation}
We can eliminate the first integral in $I_1$ by invoking the divergence theorem to write
\begin{equation}
    I_1 = \oint \bar{\eta}^j \left( 1 + \frac{1}{c^2} 2 U \right) \delta_\xi p \, d S_j - \int \left[ \partial_j \bar{\eta}^j \left( 1 + \frac{1}{c^2} 2 U \right) + \frac{1}{c^2} 3 \frac{1}{\rho^*} \bar{\eta}^j \partial_j p \right] \delta_\xi p \, d^3 x,
\end{equation}
where $dS_j$ is the outward-directed vectorial element on the star's two-surface. The stellar surface is formally defined as the radius where $p = 0$. Similar to Ref.~\cite{1965ApJ...142.1519C}, we make the assumption that $\rho$ also vanishes on the surface.%
\footnote{This assumption rules out incompressible models, but should be valid for realistic stars. The assumption can be relaxed by modifying the boundary conditions at the stellar surface.}
We then use the relation~\eqref{eq:Deltap} to find
\begin{equation}
    \frac{\delta p}{p} = - \Gamma_1 \partial_j \xi^j - \frac{1}{p} \xi^j \partial_j p - \frac{1}{c^2} 3 \Gamma_1 \left( \delta U + \frac{1}{\rho^*} \xi^j \partial_j p \right),
\end{equation}
which shows that $\delta p$ disappears on the boundary. Therefore, we arrive at
\begin{equation}
\begin{split}
    I_1 = &\int \Gamma_1 p \left( 1 + \frac{1}{c^2} 2 U \right) (\partial_j \bar{\eta}^j) (\partial_k \xi^k) \, d^3 x + \frac{1}{c^2} 3 \int \frac{1}{\rho^*} (\bar{\eta}^j \partial_j p)( \xi^k \partial_k p) \, d^3 x + \frac{1}{c^2} 3 \int  \Gamma_1 \frac{p}{\rho^*} (\bar{\eta}^j \partial_j p) (\partial_k \xi^k) \, d^3 x \\
    &+ \int \left[ 1 + \frac{1}{c^2} \left( 2 U + 3 \Gamma_1 \frac{p}{\rho^*} \right) \right] (\partial_j \bar{\eta}^j) (\xi^k \partial_k p) \, d^3 x + \frac{1}{c^2} 3 \int \Gamma_1 p (\partial_j \bar{\eta}^j) \delta_\xi U \, d^3 x.
\end{split}
\end{equation}
Here, we identify that the first two integrals satisfy the Hermitian property.

Next, we examine
\begin{equation}
    I_2 = - \int \bar{\eta}^j \left\{ \rho^* \left[ 1 + \frac{1}{c^2} \left( - U + \Pi + \frac{p}{\rho^*} \right) \right] \partial_j \delta_\xi U - \frac{1}{c^2} 3 \delta_\xi U \partial_j p \right\} \, d^3 x.
\end{equation}
To proceed, we re-write the first law~\eqref{eq:FirstLaw} using Eq.~\eqref{eq:varepsilon},
\begin{equation}
    \rho \, d\Pi = \frac{p}{\rho} \, d\rho + n_\text{b} \mu_\Delta \, dY_\text{e}.
    \label{eq:FirstLawb}
\end{equation}
Since the background is in chemical equilibrium $\mu_\Delta = 0$, this reduces to $\rho \, d\Pi = (p / \rho) \, d\rho$. Thus, again integrating by parts and using the divergence theorem, we arrive at
\begin{equation}
    I_2 = - G \int \int \frac{\overline{\delta_\eta \rho^*} \delta_\xi \rho^{* \prime}}{|x^j - x^{\prime j}|} \, d^3 x' \, d^3 x - \frac{1}{c^2} \int \left( - U + \Pi + \frac{p}{\rho^*} \right) \overline{\delta_\eta \rho^*} \delta_\xi U \, d^3 x + \frac{1}{c^2} 3 \int (\bar{\eta}^j \partial_j p) \delta_\xi U \, d^3 x.
\end{equation}
We have again used the fact that $\rho$ vanishes at the star's surface. The first term is clearly Hermitian.

Thirdly, we determine
\begin{equation}
    I_3 = - \int \bar{\eta}^j \left\{ \frac{1}{c^2} \delta_\xi \Pi + \frac{\delta_\xi \rho^*}{\rho^*} \left[ 1 + \frac{1}{c^2} \left( 2 U - \frac{p}{\rho^*} \right) \right] \right\} \partial_j p \, d^3 x.
\end{equation}
We note that Eq.~\eqref{eq:FirstLawb} also holds for Lagrangian variations, so we can state
\begin{equation}
    \rho \, \Delta \Pi = \frac{p}{\rho} \, \Delta \rho \qquad \implies \qquad \rho \, \delta \Pi = \frac{p}{\rho} \, \delta \rho,
\end{equation}
which follows from $\mu_\Delta = 0$. Hence,
\begin{equation}
\begin{split}
    I_3 &= - \int \left( 1 + \frac{1}{c^2} 2 U \right) (\bar{\eta}^j \partial_j p) \frac{\delta_\xi \rho^*}{\rho^*} \, d^3 x \\
    &= \int \frac{1}{\Gamma p} \left[ 1 + \frac{1}{c^2} \left( 2 U + 3 \Gamma \frac{p}{\rho^*} \right) \right] (\bar{\eta}^j \partial_j p) (\xi^k \partial_k p) \, d^3 x + \int \left( 1 + \frac{1}{c^2} 2 U \right) (\bar{\eta}^j \partial_j p) (\partial_k \xi^k) \, d^3 x,
\end{split}
\end{equation}
where we have used
\begin{equation}
    \frac{\delta \rho^*}{\rho^*} = - \frac{1}{\Gamma p} \xi^j \partial_j p - \partial_j \xi^j - \frac{1}{c^2} 3 \frac{1}{\rho^*} \xi^j \partial_j p
\end{equation}
and $\Gamma = d \ln p / d \ln \rho$. We note that the first term in $I_3$ is Hermitian and that the second will combine with parts of $I_1$ to form another Hermitian integral.

The fourth and final term in $\langle \eta, C \xi \rangle$ is
\begin{equation}
\begin{split}
    I_4 &= - \frac{1}{c^2} \int \bar{\eta}^j \rho^* \partial_j \delta_\xi \psi \, d^3 x \\
    &= - \frac{1}{c^2} \int \overline{\delta_\eta \rho^*} \delta_\xi \psi \, d^3 x.
\end{split}
\end{equation}
For this we use the solution~\eqref{eq:deltapsi} to write
\begin{equation}
\begin{split}
    I_4 &= - \frac{G}{c^2} \int \int \frac{\overline{\delta_\eta \rho^*}}{|x^j - x^{\prime j}|} \left[ \delta_\xi \rho^{* \prime} (- U' + \Pi') + \rho^{* \prime} \left( - \delta_\xi U' + \delta_\xi \Pi' + 3 \frac{\delta_\xi p'}{\rho^{* \prime}} \right) \right] \, d^3 x' \, d^3 x \\
    &= - \frac{1}{c^2} \int \overline{\delta_\eta U '} \left[ \delta_\xi \rho^{* \prime} (- U' + \Pi') + \rho^{* \prime} \left( - \delta_\xi U' + \delta_\xi \Pi' + 3 \frac{\delta_\xi p'}{\rho^{* \prime}} \right) \right] \, d^3 x' \\
    &= \frac{1}{c^2} \int \rho^* \overline{\delta_\eta U} \delta_\xi U \, d^3 x - \frac{1}{c^2} \int \left( - U + \Pi + \frac{p}{\rho^*} \right) \overline{\delta_\eta U} \delta_\xi \rho^* \, d^3 x + \frac{1}{c^2} 3 \int \Gamma_1 p \overline{\delta_\eta U} (\partial_j \xi^j) \, d^3 x \\
    &\quad+ \frac{1}{c^2} 3 \int \overline{\delta_\eta U} (\xi^j \partial_j p) \, d^3 x.
\end{split}
    \label{eq:I4}
\end{equation}
The first integral here is Hermitian.

Therefore, we put all the terms together to find
\begin{equation}
\begin{split}
    \langle \eta, C \xi \rangle &= I_1 + I_2 + I_3 + I_4 \\
    &= \int \Gamma_1 p \left( 1 + \frac{1}{c^2} 2 U \right) (\partial_j \bar{\eta}^j) (\partial_k \xi^k) \, d^3 x + \int \frac{1}{\Gamma p} \left[ 1 + \frac{1}{c^2} \left( 2 U + 6 \Gamma \frac{p}{\rho^*} \right) \right] (\bar{\eta}^j \partial_j p) (\xi^k \partial_k p) \, d^3 x \\
    &\quad+ \int \left[ 1 + \frac{1}{c^2} \left( 2 U + 3 \Gamma_1 \frac{p}{\rho^*} \right) \right] (\bar{\eta}^j \partial_j p \partial_k \xi^k + \partial_j \bar{\eta}^j \xi^k \partial_k p) \, d^3 x + \frac{1}{c^2} 3 \int \Gamma_1 p (\partial_j \bar{\eta}^j \delta_\xi U + \overline{\delta_\eta U} \partial_j \xi^j) \, d^3 x \\
    &\quad- G \int \int \frac{\overline{\delta_\eta \rho^*} \delta_\xi \rho^{* \prime}}{|x^j - x^{\prime j}|} \, d^3 x' \, d^3 x - \frac{1}{c^2} \int \left( - U + \Pi + \frac{p}{\rho^*} \right) (\overline{\delta_\eta \rho^*} \delta_\xi U + \overline{\delta_\eta U} \delta_\xi \rho^*) \, d^3 x \\
    &\quad+ \frac{1}{c^2} 3 \int (\bar{\eta}^j \partial_j p \delta_\xi U + \overline{\delta_\eta U} \xi^j \partial_j p) \, d^3 x + \frac{1}{c^2} \int \rho^* \overline{\delta_\eta U} \delta_\xi U \, d^3 x \\
    &= \overline{\langle \xi, C \eta \rangle},
\end{split}
\end{equation}
which is manifestly Hermitian. This accords with \citet{1965ApJ...142.1519C}. Chandrasekhar's interest in demonstrating the self-adjoint nature of the perturbation problem lay in deriving a variational basis for the eigenfrequencies of the mode solutions. Our aim is different in that we intend to use the self-adjointness to motivate a mode-sum representation in pN theory.

\subsection{Orthogonality and the mode-sum representation}

Since the operators $A_{j k}$ and $C_{j k}$ are both Hermitian, the equation of motion~\eqref{eqs:PerturbedEuler} follows from a variational principle with action
\begin{equation}
    S = \int \mathcal{L} \, d^3 x = \frac{1}{2} \int (\partial_t \xi^j A_{j k} \partial_t \xi^k - \xi^j C_{j k} \xi^k) \, d^3 x,
\end{equation}
which defines the Lagrangian density $\mathcal{L}$ of the perturbed system. It is straightforward to verify that a variation of $\mathcal{L}$ leads to \eqref{eqs:PerturbedEuler}.

The Lagrangian system is associated with a symplectic structure $W$ from which we can derive conserved quantities associated with the symmetries of the equilibrium (see Ref.~\cite{1978ApJ...221..937F} for more detail). The symplectic structure is given by
\begin{equation}
    W(\eta, \xi) = \langle \eta, A \partial_t \xi \rangle - \langle A \partial_t \eta, \xi \rangle.
\end{equation}
It has the useful property that it is always conserved
\begin{equation}
    \frac{d}{dt} W(\eta, \xi) = 0,
\end{equation}
the proof of which relies on the static nature of the background, as well as the symmetry properties we have just demonstrated. Using the symplectic structure, we define the canonical energy of a perturbation as
\begin{equation}
    E_\text{c} = \frac{1}{2} W(\partial_t \xi, \xi) = \frac{1}{2} \langle \partial_t \xi, A \partial_t \xi \rangle + \frac{1}{2} \langle \xi, C \xi \rangle.
\end{equation}
The canonical energy is associated with the time-symmetry of the perturbation equations; since the equilibrium is time-independent, $\partial_t \xi^j$ satisfies the equation of motion when $\xi^j$ is a solution. It is a simple exercise to demonstrate that the canonical energy is conserved.

Our main interest here lies in taking advantage of the Hermitian nature of the perturbation problem to form a mode-expansion for the tidal response in a binary system. The mode-sum representation has been well explored in Newtonian gravity \cite{1979RSPSA.368..389D,2001PhRvD..65b4001S} and proved to be a useful tool for studying the dynamical tide of neutron stars \cite{1994ApJ...426..688R,1994MNRAS.270..611L,2020PhRvD.101h3001A}. However, such an expansion is not formally permitted in general relativity. Therefore, it may be useful for relativistic contexts to assess whether progress can be made in the pN approximation. Thus, we consider two normal-mode solutions
\begin{equation}
    \xi^j(t, x^k) = \xi_\alpha^j(x^k) e^{i \omega_\alpha t}, \qquad \eta^j(t, x^k) = \xi_\beta^j(x^k) e^{i \omega_\beta t}
\end{equation}
to the equation of motion~\eqref{eqs:PerturbedEuler}, where $\omega_\alpha$ and $\omega_\beta$ are eigenfrequencies with corresponding eigenfunctions $\xi_\alpha^j$ and $\xi_\beta^j$. (Recall that we are working to first pN order and treating the stellar material as a perfect fluid. There is therefore no dissipation and the mode eigenfrequencies are real.) The normal modes are solutions to the eigenvalue problem
\begin{equation}
    0 = - \omega_\alpha^2 A_{j k} \xi_\alpha^k + C_{j k} \xi_\alpha^k.
\end{equation}
We examine the symplectic product between these two solutions to find
\begin{equation}
    \frac{d}{dt} W(\eta, \xi) = - (\omega_\alpha^2 - \omega_\beta^2) \langle \xi_\beta, A \xi_\alpha \rangle e^{i (\omega_\alpha - \omega_\beta) t},
\end{equation}
which leads to the orthogonality condition
\begin{equation}
    \langle \xi_\beta, A \xi_\alpha \rangle = \mathcal E_\alpha \delta_{\alpha \beta},
    \label{eq:Orthogonal}
\end{equation}
where $\mathcal{E}_\alpha$ is a normalisation factor. This demonstrates that distinct mode solutions are orthogonal to each other with respect to the inner product in Eq.~\eqref{eq:Orthogonal}. Equipped with this result, we may decompose a generic perturbation of the star as a sum over all the modes. We will carry this out in a moment.

For a given mode, the canonical energy becomes
\begin{equation}
    E_\text{c} = \omega_\alpha^2 \mathcal{E}_\alpha
\end{equation}
and we can identify
\begin{equation}
    \mathcal{E}_\alpha = \langle \xi_\alpha, A \xi_\alpha \rangle = \int \bar{\xi}_\alpha^j \rho^* \left\{ \left[ 1 + \frac{1}{c^2} \left( 3 U + \Pi + \frac{p}{\rho^*} \right) \right] \xi_{\alpha j} - \frac{1}{c^2} \left( 4 \delta_{\xi_\alpha} V_j + \frac{1}{2} \partial_j \delta_{\xi_\alpha} X \right) \right\} \, d^3 x.
\end{equation}
There is an important subtlety with regards to the canonical energy that is worth commenting on. In general, $E_\text{c}$ is \textit{not} the full energy of the perturbation \cite{1978ApJ...221..937F}. However, in the Newtonian theory, the canonical energy does in fact coincide with the physical energy when the star is non-rotating. If this is also the case in the pN approximation remains to be established. The star's change in energy due to a perturbation is ultimately a second-order perturbative quantity. Therefore, to determine the energy in pN theory, it would be necessary to conduct a second-order  computation. As this is beyond the scope of the present paper, it is left for future work.

We now demonstrate how a generic perturbation can be decomposed using the normal modes in pN theory. Suppose that the star is deformed by an external force density $f_j$, such that the equation of motion becomes
\begin{equation}
    f_j = A_{j k} \partial_t^2 \zeta^k + C_{j k} \zeta^k,
    \label{eq:Source}
\end{equation}
where $\zeta^j$ is the Lagrangian displacement in response to $f_j$. The solution to this problem can be obtained by decomposing the stellar response as
\begin{equation}
    \zeta^j(t, x^k) = \sum_\alpha a_\alpha(t) \xi_\alpha^j(x^k),
\end{equation}
where $a_\alpha(t)$ is an amplitude and the summation over $\alpha$ includes all the distinct mode solutions (see, \textit{e.g.}, Ref.~\cite{2020PhRvD.101h3001A}). Therefore, the equation of motion~\eqref{eq:Source} reduces to that of a forced harmonic oscillator
\begin{equation}
    \frac{d^2 a_\alpha}{dt^2} + \omega_\alpha^2 a_\alpha = \frac{Q_\alpha}{\mathcal{E}_\alpha},
    \label{eq:Oscillator}
\end{equation}
where
\begin{equation}
    Q_\alpha(t) = \langle \xi_\alpha, f(t) \rangle
\end{equation}
is the \textit{overlap integral} between the mode and the deforming force, and the orthogonality condition~\eqref{eq:Orthogonal} has been used. This result has the same form as in Newtonian gravity---where the important difference is due to the fact that $A_{j k}$ and $C_{j k}$ here include pN corrections---and shows how normal modes computed in the pN approximation form a complete basis. That this decomposition is possible simplifies the full, partial differential equations problem of Eq.~\eqref{eq:Source} to a one-dimensional, ordinary differential equation given by Eq.~\eqref{eq:Oscillator}.

\subsection{Uniqueness of the formulation}

We have demonstrated how the perturbation equations in the form of \eqref{eqs:PerturbedEuler} are Hermitian and therefore admit mode solutions that form a complete basis. We will now examine the uniqueness of this formulation.

We define two new operators $\mathcal{A}_{j k}$ and $\mathcal{C}_{j k}$ by multiplying \eqref{eqs:PerturbedEuler} by $(1 + F / c^2)$, where $F$ is a function to be determined. We do this to see if there are other non-trivial solutions that are Hermitian. Showing that there are none will demonstrate the form~\eqref{eqs:PerturbedEuler} is unique and leads to a well-defined canonical energy and mode-orthogonality condition in pN theory. Since we are working to first pN order, we ignore terms of $O(c^{-4})$. With the transformation, the equation of motion becomes
\begin{subequations}
\begin{equation}
    0 = \mathcal{A}_{j k} \partial_t^2 \xi^k + \mathcal{C}_{j k} \xi^k,
\end{equation}
where
\begin{align}
    \mathcal{A}_{j k} \partial_t^2 \xi^k &= \rho^* \left\{ \left[ 1 + \frac{1}{c^2} \left( F + 3 U + \Pi + \frac{p}{\rho^*} \right) \right] \partial_t^2 \xi_j - \frac{1}{c^2} \left( 4 \partial_t^2 \delta V_j + \frac{1}{2} \partial_t^2 \partial_j \delta X \right) \right\}, \\
\begin{split}
    \mathcal{C}_{j k} \xi^k &= \left[ 1 + \frac{1}{c^2} (F + 2 U) \right] \partial_j \delta p - \rho^* \left[ 1 + \frac{1}{c^2} \left( F - U + \Pi + \frac{p}{\rho^*} \right) \right] \partial_j \delta U - \frac{1}{c^2} \left( - 3 \delta U + \delta \Pi + \frac{\delta p}{\rho^*} \right) \partial_j p \\
    &\qquad- \frac{\delta \rho^*}{\rho^*} \left[ 1 + \frac{1}{c^2} \left( F + 2 U - \frac{p}{\rho^*} \right) \right] \partial_j p - \frac{1}{c^2} \rho^* \partial_j \delta \psi.
\end{split}
\end{align}
\end{subequations}
We can immediately see that $\mathcal{A}_{j k}$ satisfies the same symmetry property as $A_{j k}$,
\begin{equation}
    \langle \eta, \mathcal{A} \xi \rangle = \overline{\langle \xi, \mathcal{A} \eta \rangle},
\end{equation}
irrespective of the form of $F$. After some algebra and various manipulations, for $\mathcal{C}_{j k}$, we find
\begin{equation}
\begin{split}
    \langle \eta, \mathcal{C} \xi \rangle &= \int \Gamma_1 p \left[ 1 + \frac{1}{c^2} (F + 2 U) \right] (\partial_j \bar{\eta}^j) (\partial_k \xi^k) \, d^3 x + \int \frac{1}{\Gamma p} \left[ 1 + \frac{1}{c^2} \left( F + 2 U + 6 \Gamma \frac{p}{\rho^*} \right) \right] (\bar{\eta}^j \partial_j p) (\xi^k \partial_k p) \, d^3 x \\
    &\quad+ \int \left[ 1 + \frac{1}{c^2} \left( F + 2 U + 3 \Gamma_1 \frac{p}{\rho^*} \right) \right] (\bar{\eta}^j \partial_j p \partial_k \xi^k + \partial_j \bar{\eta}^j \xi^k \partial_k p) \, d^3 x \\
    &\quad+ \frac{1}{c^2} 3 \int \Gamma_1 p (\partial_j \bar{\eta}^j \delta_\xi U + \overline{\delta_\eta U} \partial_j \xi^j) \, d^3 x - G \int \int \frac{\overline{\delta_\eta \rho^*} \delta_\xi \rho^{* \prime}}{|x^j - x^{\prime j}|} \, d^3 x' \, d^3 x \\
    &\quad- \frac{1}{c^2} \int \left( - U + \Pi + \frac{p}{\rho^*} \right) (\overline{\delta_\eta \rho^*} \delta_\xi U + \overline{\delta_\eta U} \delta_\xi \rho^*) \, d^3 x + \frac{1}{c^2} 3 \int (\bar{\eta}^j \partial_j p \delta_\xi U + \overline{\delta_\eta U} \xi^j \partial_j p) \, d^3 x  \\
    &\quad+ \frac{1}{c^2} \int \rho^* \overline{\delta_\eta U} \delta_\xi U \, d^3 x + \frac{1}{c^2} \int (\bar{\eta}^j \partial_j F) (\Gamma_1 p \partial_k \xi^k + \xi^k \partial_k p + \rho^* \delta_\xi U) \, d^3 x - \frac{1}{c^2} \int F \overline{\delta_\eta \rho^*} \delta_\xi U \, d^3 x.
\end{split}
\end{equation}
It is clear that all the terms are Hermitian, except for the final two integrals, which require a more careful examination. Collecting them together, we have
\begin{equation}
    J = \frac{1}{c^2} \int (\bar{\eta}^j \partial_j F) (\Gamma_1 p \partial_k \xi^k + \xi^k \partial_k p + \rho^* \delta_\xi U) \, d^3 x - \frac{1}{c^2} \int F \overline{\delta_\eta \rho^*} \delta_\xi U \, d^3 x.
    \label{eq:J}
\end{equation}
Evidently, if $F = 0$, $J$ vanishes and we retain the earlier formulation~\eqref{eqs:PerturbedEuler}. However, we want to know if there are non-trivial solutions.

We take advantage of the Poisson equation $\nabla^2 \delta U = - 4 \pi G \delta \rho^*$ [\textit{cf.}, Eq.~\eqref{eq:U}] to manipulate Eq.~\eqref{eq:J} as
\begin{equation}
    J = \frac{1}{c^2} \int (\bar{\eta}^j \partial_j F) (\Gamma_1 p \partial_k \xi^k + \xi^k \partial_k p + \rho^* \delta_\xi U) \, d^3 x + \frac{1}{c^2} \frac{1}{4 \pi G} \oint F \partial^j \overline{\delta_\eta U} \delta_\xi U \, dS_j - \frac{1}{c^2} \frac{1}{4 \pi G} \int \partial^j \overline{\delta_\eta U} \partial_j (F \delta_\xi U) \, d^3 x.
    \label{eq:J2}
\end{equation}
At this point, the integration domain becomes important. The integrand in the surface integral is not bound by the stellar matter since the linearised potentials exist in the vacuum.

Suppose the integration extends over all three-space. The surface integral then immediately vanishes as the perturbed potentials fall off at infinity (with the caveat that $F$ must grow slower than $\partial_r \overline{\delta_\eta U} \delta_\xi U$ decays). This leads to an obvious alternative to $F = 0$: The operator $\mathcal{C}_{j k}$ is Hermitian when $\partial_j F = 0$, which is satisfied for
\begin{equation}
    F \propto \left( - U + \Pi + \frac{p}{\rho^*} \right).
\end{equation}
This implies that there is a whole class of formulations of the equations of motion that obey the necessary symmetry properties and naturally raises questions about the uniqueness of the canonical energy $E_\text{c}$.

Fortunately, there is a vital flaw in this argument: the integration does not penetrate the exterior. As we have discussed, the volume integration is inextricably tethered to the stellar fluid due to the compact support of the integrands. This is also the case here, since it is $\overline{\delta_\eta \rho^*}$ that initially appeared in Eq.~\eqref{eq:J}. That this is not immediately apparent in Eq.~\eqref{eq:J2} is because we made use of a Poisson equation to replace $\overline{\delta_\eta \rho^*}$ in favour of $\nabla^2 \overline{\delta_\eta U}$. But the Poisson equation is still tied to the stellar volume, since it vanishes outside the star. Hence, the surface integral in Eq.~\eqref{eq:J2} takes place at the stellar boundary and is not generically zero.

All of this means that the only way to make the perturbation equations satisfy the Hermitian symmetries is to demand that $F = 0$, proving that the form of the equation of motion~\eqref{eqs:PerturbedEuler} is unique and defines the operators $A_{j k}$ and $C_{j k}$ at 1pN. This is important since, when we linearised the pN Euler equation~\eqref{eq:EulerFull}, we chose to multiply the result by a factor of $[1 + (3 U + \Pi + p / \rho^*) / c^2]$ to obtain the same result as Ref.~\cite{1965ApJ...142.1519C}. As we have shown, without this pN correction, the resultant equation of motion would not  satisfy the necessary symmetry properties.

\section{Conclusions}

The pN approximation assumes that the physical context involves slow motion and weak gravitational fields. Even though this is not the appropriate regime for relativistic neutron stars, there are problems where the modelling is technically challenging and progress beyond Newtonian models can be made. A contemporary example of this is the dynamical tide of binary neutron stars, which is relevant for gravitational-wave astronomy. Given this, we have considered using pN theory to describe neutron-star dynamics.

In perturbation analysis, it is common to decompose the relevant quantities using a small parameter and then calculate at each successive order in the small parameter. Although there exists a small parameter implicit within pN theory, we showed how there is ambiguity in the thermodynamics treatment that causes issues for a strict order-by-order expansion.

We went on to present the standard derivation of the pN Euler equation from the relativistic fluid equations using the isotropic pN metric and considered linear perturbations of a non-rotating star to arrive at the pN perturbation equations. We demonstrated the expected result that the perturbed equation of motion~\eqref{eqs:PerturbedEuler} is Hermitian. We showed that the equation of motion follows from a fundamental Lagrangian, which implies that---just as in Newtonian gravity---the normal modes up to 1pN are complete. Therefore, a generic perturbation in pN theory can be described as a mode-sum. We have demonstrated how the mode-sum representation reduces the problem of a star perturbed by an external force to that of a forced harmonic oscillator.

In the pN approximation, one has freedom to manipulate the expressions by discarding higher order corrections. Therefore, we examined the uniqueness of the form~\eqref{eqs:PerturbedEuler} in being Hermitian. Paying careful attention to the integration domain, we argue that there is no pN correction one can make to the equation of motion that satisfies the relevant Hermitian symmetries. This means that the perturbation equation of the form~\eqref{eqs:PerturbedEuler} is unique and follows from a well-defined Lagrangian system that possesses a well-defined canonical energy at 1pN. Furthermore, for oscillation modes this leads to a unique definition of the orthogonality condition necessary for the mode-sum representation.

The motivation behind this work was to set the stage for describing dynamical tides in pN theory. Hence, we focused on the formal demonstration that the perturbation equations in pN derive from a fundamental Lagrangian system. This, in turn, implies that oscillation modes in pN theory may be used as a complete basis for the tidal response of a star. There are several interesting directions for future work. For instance:

\begin{enumerate}
    \item The obvious next step will be to calculate modes of oscillation and  explore the extent to which pN theory can usefully be applied to the dynamical tides of neutron stars. This work is already well under way and we expect to report on it soon \cite{pn_modes}.
    \item Further developing the pN perturbation formalism we have set out in this paper, another promising direction for future study would be to extend the approach to rotating stars. We know from Newtonian gravity that this introduces an extra term that is linear in $\partial_t \xi^j$ to the equation of motion \cite{1978ApJ...221..937F}. In addition, fluid motion in the background generates the post-Newtonian vector potential $U_j$, introducing \textit{gravito-magnetic} effects into the problem (see, \textit{e.g.}, Ref.~\cite{2020PhRvD.101j4028P}). The impact of these features would be interesting to explore.
    \item Along a similar vein, computing the modes of rotating stars remains a formidable task in general relativity and it may be valuable to tackle the calculation in the pN approximation. (This was, in fact, the motivation behind Refs.~\cite{1991ApJ...374..248C,1992ApJ...385..630C}.) It is at least conceivable that a calculation in the pN approximation will be tractable and lead to more accurate results than in the Newtonian theory. Furthermore, in relativity, the determination of the astrophysically interesting \textit{r}-modes leads to a singular eigenvalue problem that we are (at present) unable to solve \cite{1998MNRAS.293...49K}. It seems plausible that approaching this problem from a pN perspective may provide useful hints (see, \textit{e.g.}, Ref.~\cite{2000PhRvD..63b4019L}).
    \item From a formal point of view, extending the formalism to second order, both for perturbations and the pN approximation, would be highly beneficial. This extension would be relevant for calculating the physical energy of the perturbations and comparing it to the canonical energy. Additionally, we suspect that the oscillation modes remain complete up to 2pN, since dissipation due to gravitational-wave emission arises at 2.5pN. Verifying this conjecture would be an interesting study in its own right and would set the stage for exploring the perturbation problem at 2.5pN.
\end{enumerate}

\begin{acknowledgements}
    F.G. acknowledges funding from the European Union’s Horizon Europe research and innovation programme under the Marie Sk{\l}odowska-Curie grant agreement No.~101151301.
    N.A. acknowledges support from STFC via grant No.~ST/R00045X/1.
\end{acknowledgements}

\bibliography{refs.bib}

\end{document}